\documentclass[a4paper,12pt]{article}
\usepackage{graphicx}
\usepackage{amsthm}
\usepackage{amsfonts}
\usepackage{amsmath,latexsym}
\usepackage{amssymb}
\usepackage{hyperref}
\usepackage{tabularx}
\newcolumntype{M}{>{$}c<{$}}
\usepackage[matrix,arrow,curve]{xy}

\numberwithin{equation}{section} \numberwithin{figure}{section}
\numberwithin{table}{section}

\def\papertitlepage{\baselineskip 3.5ex\thispagestyle{empty}}
\def\Title#1{\baselineskip 1cm \vspace{1.5cm}%
  \begin{center}{\Large\bf #1}\end{center}\vspace{0.5cm}}
\def\Authors#1{\begin{center}\renewcommand{\thefootnote}{\fnsymbol{footnote}}{\it #1}\end{center}}
\def\Abstract{\vspace{1.0cm}%
  \begin{center}{\large\bf Abstract}\end{center}}

\renewenvironment{thebibliography}{\pagebreak[3]\par\vspace{0.6em}
\begin{flushleft}{\large \bf References}\end{flushleft}
\vspace{-1.0em}

\begin{enumerate}\if@twocolumn\baselineskip=0.6em\itemsep -0.2em
\else\itemsep -0.2em\fi\labelsep 0.1em}{\end{enumerate} }

\begin{document}
{\papertitlepage \vspace*{0cm} {\hfill
\begin{minipage}{4.2cm}
CCNH-UFABC 2016\par\noindent April, 2016
\end{minipage}}
\Title{A singular one-parameter family of solutions in cubic superstring field theory}
\Authors{{\sc E.~Aldo~Arroyo${}$\footnote{\tt
aldo.arroyo@ufabc.edu.br}}
\\
Centro de Ci\^{e}ncias Naturais e Humanas, Universidade Federal do ABC \\[-2ex]
Santo Andr\'{e}, 09210-170 S\~{a}o Paulo, SP, Brazil ${}$ }
} 

\vskip-\baselineskip
{\baselineskip .5cm \Abstract Performing a gauge transformation of
a simple identity-like solution of superstring field theory, we
construct a one-parameter family of solutions, and by evaluating
the energy associated to this family, we show that for most of the
values of the parameter the solution represents the tachyon
vacuum, except for two isolated singular points where the solution
becomes the perturbative vacuum and the half brane solution.
 }
\newpage
\setcounter{footnote}{0}
\tableofcontents

\section{Introduction}
It is known that the analytic solutions for tachyon condensation
\cite{Schnabl:2005gv,Erler:2009uj,Okawa:2006vm} in open bosonic
string field theory \cite{Witten:1985cc} as well as the ones
\cite{Erler:2007xt,Aref'eva:2008ad,Gorbachev:2010zz} in cubic
superstring field theory \cite{Arefeva:1989cp} are formally gauge
equivalent to identity based solutions
\cite{Arroyo:2010fq,Zeze:2010sr,Arroyo:2010sy,Arefeva:2010yd,Erler:2012qn}.
Identity based solutions are constructed as a product of certain
linear combination of ghost number one operators with the identity
string field
\cite{Kishimoto:2014lua,Kishimoto:2009nd,Inatomi:2011an}.

Although identity based solutions are pathological solutions in the sense that they bring ambiguous analytic result for the value of the energy
\cite{Erler:2012dz}, by performing a gauge transformation over these solutions, it is possible to construct well behaved solutions. For instance, in
reference \cite{Zeze:2010sr}, a one-parameter family of solutions has been found which interpolates between an identity based solution and the
Erler-Schnabl's tachyon vacuum solution \cite{Erler:2009uj}. This result has been extended for the case of cubic superstring field theory
\cite{Arroyo:2010sy}, namely, a one-parameter family of solutions has been found which interpolates between an identity based solution and the Gorbachev's
tachyon vacuum solution \cite{Gorbachev:2010zz}.

Motivated by the above results, and the recently discovered
Erler's half brane solution \cite{Erler:2010pr} in cubic
superstring field theory, in this paper, starting with the
identity based solution
\cite{Arroyo:2010fq,Arefeva:2010yd,AldoArroyo:2012if,Arroyo:2013pha}
\begin{align}
\label{Iden1Intro}  \widehat{\Phi}_I = \Big(  (c+B \gamma^2)(1-K)
\Big) \otimes \sigma_3,
\end{align}
by performing a gauge transformation of $\widehat{\Phi}_I $, we
study the construction of the following one-parameter family of
solutions
\begin{eqnarray}
\label{SolphiIntro} \widehat{\Phi}_{\lambda} = \Phi_{1,\lambda}
\otimes \sigma_3 + \Phi_{2,\lambda} \otimes i \sigma_2,
\end{eqnarray}
where the string fields $\Phi_{1,\lambda}$ and $\Phi_{2,\lambda}$
are given by
\begin{eqnarray}
\label{solphi3In} \Phi_{1,\lambda} &=& Q(Bc)f(K,\lambda) + \lambda(2\lambda-1)c f(K,\lambda) + 4i\lambda(1-\lambda) c GB c G \widetilde{f}(K,\lambda) ,\\
\label{solphi4In} \Phi_{2,\lambda} &=& Q(Bc) G
\widetilde{f}(K,\lambda) + \lambda(2\lambda-1)c G
\widetilde{f}(K,\lambda) + 4i \lambda(1-\lambda) c G B c
f(K,\lambda),
\end{eqnarray}
with $f(K,\lambda)$ and $\widetilde{f}(K,\lambda)$ being functions
of $K$\footnote{The $K$ field is an element of the so-called $KBc$
subalgebra introduced in the references
\cite{Okawa:2006vm,Erler:2006hw,Erler:2006ww,Schnabl:2010tb}.} and
the parameter $\lambda$
\begin{eqnarray}
\label{gaugeF1In} f(K,\lambda) &=&  \frac{\lambda^2 (1-2 \lambda
)^2+\left(16 \lambda^3-32 \lambda^2+18 \lambda -1\right) \lambda
\, K }{\lambda ^2(1-2
   \lambda )^2 +2 \lambda  \left(8 \lambda ^3-16 \lambda ^2+10 \lambda -1\right) K+K^2} \; , \\
\label{gaugeF2In} \widetilde{f}(K,\lambda) &=&   \frac{4 i
(1-\lambda) \lambda \, K}{\lambda ^2(1-2
   \lambda )^2 +2 \lambda  \left(8 \lambda ^3-16 \lambda ^2+10 \lambda -1\right) K+K^2} \;
   .
\end{eqnarray}

Moreover, by explicit and detailed computation of the normalized
value of the energy
\begin{eqnarray}
\label{NorV1Intro} E(\widehat{\Phi}_{\lambda}) = \frac{\pi^2}{3}
\Big[\langle Y_{-2} \Phi_{1,\lambda} Q \Phi_{1,\lambda} \rangle +
\langle Y_{-2} \Phi_{2,\lambda} Q \Phi_{2,\lambda} \rangle \Big]
\end{eqnarray}
associated to the solution $\widehat{\Phi}_{\lambda}$, we obtain
\begin{align}
\label{FinalResult1Intro} E(\widehat{\Phi}_{\lambda}) =
\begin{cases}
      0, & \lambda = 0 \;,\;\; \text{Perturbative Vacuum Solution,} \\
      -1/2, & \lambda = 1/2 \;,\;\; \text{Half Brane Solution,} \\
      -1, & \big(\lambda < 0\big) \vee \big(\kappa \leq
\lambda <\frac{1}{2}\big) \vee \big(\lambda >\frac{1}{2}\big)
\;,\;\; \text{Tachyon Vacuum Solution,}
   \end{cases}
\end{align}
where $\kappa$ is a numerical constant defined as
\begin{eqnarray}
\label{kappa1Intro} \kappa = \frac{2}{3}-\frac{1}{6}
\left(\frac{25}{2}+\frac{3}{2} \sqrt{69}\right)^{1/3}-\frac{1}{6}
\left(\frac{25}{2}-\frac{3}{2} \sqrt{69}\right)^{1/3} \approx
0.122561.
\end{eqnarray}

Note that for most of the values of the parameter $\lambda$, the solution represents the tachyon vacuum, while the two isolated points $\lambda=0$ and
$\lambda=1/2$ correspond to the perturbative vacuum and the half brane solution respectively.

We expect that the construction of a one-parameter family of solutions using identity based solutions, in cubic superstring field theory, will provide us
with relevant tools to analyze other important solutions, such as the multibrane solutions \cite{AldoArroyo:2012if,Arroyo:2013pha}, and the recently
proposed Erler's analytic solution for tachyon condensation in Berkovits non-polynomial open superstring field theory \cite{Erler:2013wda}. Since the
algebraic structure of Berkovits theory \cite{Berkovits:1995ab} is similar to the cubic superstring field theory, the results of our work can be naturally
extended, however, the presence of a non-polynomial action in Berkovits theory will bring us challenges in the search of new solutions.

This paper is organized as follows. In section 2, we review the
modified cubic superstring field theory and introduce some
notations and conventions.  Since the explicit form of our
one-parameter family of solutions is expressed in terms of
elements of the $GKBc\gamma$ algebra, in section 3, we study in
detail this algebra. In section 4, by performing a gauge
transformation of an identity based solution, we show the
construction of the one-parameter family of solutions. In section
5, we analyze correlation functions involving the $G$ field and as
a pedagogical application of these correlators, we show the
computation of the energy for the half brane solution. In section
6, we evaluate the energy associated to the one-parameter family
of solutions. In section 7, a summary and further directions of
exploration are given.

\section{Modified cubic superstring field theory, notations and conventions}
The action of the modified cubic superstring field theory which
takes into account the $GSO(+)$ and $GSO(-)$ sectors is given by
\cite{Arefeva:1989cp}
\begin{eqnarray}
\label{action1} S = -\frac{1}{g^2} \Big[\frac{1}{2} \langle Y_{-2}
\Phi_1 Q \Phi_1 \rangle + \frac{1}{3} \langle Y_{-2} \Phi_1 \Phi_1
\Phi_1 \rangle + \frac{1}{2} \langle Y_{-2} \Phi_2 Q \Phi_2
\rangle - \langle Y_{-2} \Phi_1  \Phi_2 \Phi_2 \rangle  \Big],
\end{eqnarray}
where $Q$ is the BRST operator of the open Neveu-Schwarz
superstring theory. The operator $Y_{-2}$ is inserted at the open
string midpoint and is written as the product of two inverse
picture changing operators $Y_{-2}=Y(i)Y(-i)$, where
$Y(z)=-\partial \xi e^{-2 \phi} c(z)$. The ghost number one string
fields $\Phi_1$ and $\Phi_2$ belong to the $GSO(+)$ and $GSO(-)$
sectors, and are Grassman odd and Grassman even respectively.

Varying the action (\ref{action1}) with respect to the string
fields $\Phi_1$ and $\Phi_2$ yields the following equations of
motion \cite{Arefeva:1988nn}
\begin{eqnarray}
\label{Eqm1} Q \Phi_1 + \Phi_1*\Phi_1 - \Phi_2*\Phi_2 &=& 0 , \\
\label{Eqm2} Q \Phi_2 + \Phi_1*\Phi_2 - \Phi_2*\Phi_1 &=& 0.
\end{eqnarray}
Regarding to the star product, we are going to use the left handed
convention of \cite{Schnabl:2005gv,Erler:2009uj}. There are other
sources which use the right handed convention
\cite{Okawa:2006vm,Berkovits:2000hf}, for details related to the
connection between these two conventions see reference
\cite{Erler:2010pr}.

Using the equations of motion (\ref{Eqm1}), (\ref{Eqm2}) and the
cyclicity relation
\begin{eqnarray}
\label{Rela1} \langle Y_{-2} \Phi_1 \Phi_2 \Phi_2  \rangle = -
\langle Y_{-2} \Phi_2 \Phi_1 \Phi_2 \rangle = \langle Y_{-2}
\Phi_2 \Phi_2 \Phi_1 \rangle,
\end{eqnarray}
where an additional minus sign arises due to the fact that
$\Phi_2$ belongs to the $GSO(-)$ sector\footnote{Since a string
field belonging to the $GSO(-)$ sector has half-integer conformal
weight, $\Phi_2$ changes its sign under the conformal
transformation $\mathcal{R}_{2\pi}$ representing the $2\pi$
rotation of the unit disk \cite{Ohmori:2003vq}.}, we can write the
action (\ref{action1}) as
\begin{eqnarray}
\label{action2} S = -\frac{1}{6 g^2} \Big[\langle Y_{-2} \Phi_1 Q
\Phi_1 \rangle + \langle Y_{-2} \Phi_2 Q \Phi_2 \rangle  \Big].
\end{eqnarray}

Since $\Phi_2$ has opposite Grassmannality as compared to the $GSO(+)$ string field $\Phi_1$, it seems that they fail to obey common algebraic relations.
This problem can be resolved by attaching the $2 \times 2$ internal Chan-Paton matrices to the string fields and the operator insertions as
\cite{Berkovits:2000hf,Arefeva:2002mb}
\begin{align}
\label{chanfield1}  \widehat{Q} &= Q \otimes \sigma_3 , \;\;\;\;
\widehat{Y}_{-2} = Y_{-2} \otimes \sigma_3, \\
\label{chanfield2}  \widehat{\Phi} &= \Phi_1 \otimes \sigma_3 +
\Phi_2 \otimes i \sigma_2.
\end{align}
Using these definitions, the action (\ref{action1}) can be written
in a compact way
\begin{eqnarray}
\label{action3} S = -\frac{1}{2 g^2} \text{Tr}\Big[  \frac{1}{2}
\langle \widehat{Y}_{-2} \widehat{\Phi} \widehat{Q} \widehat{\Phi}
\rangle + \frac{1}{3} \langle \widehat{Y}_{-2} \widehat{\Phi}
\widehat{\Phi} \widehat{\Phi} \rangle \Big],
\end{eqnarray}
and the equations of motion (\ref{Eqm1}) and (\ref{Eqm2}) are
reduced to a single equation
\begin{eqnarray}
\label{EqmC1} \widehat{Q }\widehat{\Phi} + \widehat{\Phi}
\widehat{\Phi} = 0 .
\end{eqnarray}

For a given ghost number zero string field $\widehat{U}= U_1
\otimes \mathbb{I} + U_2 \otimes \sigma_1$, we can construct a
gauge transformation of the string field $\widehat{\Phi}$ as
follows
\begin{eqnarray}
\label{Gauge1} \widehat{\Psi} = \widehat{U} (\widehat{Q} +
\widehat{\Phi}) \widehat{U}^{-1} .
\end{eqnarray}
It turns out that the action (\ref{action3}) is invariant under
this gauge transformation (\ref{Gauge1}). If $\widehat{\Phi}$ is a
solution of the equation of motion (\ref{EqmC1}) then a string
field $\widehat{\Psi}$, related to $\widehat{\Phi}$ by means of
the equation (\ref{Gauge1}), is also a solution.

In order to find analytic solutions of the equation of motion
(\ref{EqmC1}), we can employ the prescription studied in reference
\cite{Arroyo:2010fq}, namely, (i) find a simplest identity based
solution of the equation of motion\footnote{Although the identity
based solution formally satisfies the equation of motion
(\ref{EqmC1}), it is a pathological solution in the sense that it
brings ambiguous analytic result for the value of the energy
\cite{Arroyo:2010fq,Kishimoto:2014lua,Kishimoto:2009nd,Inatomi:2011an}.},
(ii) perform a gauge transformation over this identity based
solution such that the resulting string field, consistently,
represents a well behaved solution
\cite{Arroyo:2010sy,Zeze:2010sr}.

In this paper, following the above procedures, we are going to construct a one-parameter family of solutions $\widehat{\Phi}_{\lambda}$ and evaluate the
energy associated to these solutions. It turns out that, depending on the value of the parameter $\lambda$, the solutions $\widehat{\Phi}_{\lambda}$
describe three distinct gauge orbits corresponding to the perturbative vacuum, the half brane and the tachyon vacuum solution. Before deriving the explicit
form of the solution $\widehat{\Phi}_{\lambda}$, in the next section we will introduce the so-called $GKBc\gamma$ algebra.

\section{The $GKBc\gamma$ algebra, definitions and star products}
The $GKBc\gamma$ algebra is an extension of the well known
$KBc\gamma$ algebra \cite{Erler:2006hw,Erler:2006ww,Erler:2007xt}.
Essentially, we add the new element $G$ to the $KBc\gamma$
algebra. This string field $G$ lives in the $GSO(-)$ sector, and
is related to the worldsheet supercurrent $G(z)$
\cite{Erler:2010pr}.

To derive some identities involving the star product of the basic
string fields $G$, $K$, $B$, $c$ and $\gamma$ together with the
action of the BRST operator $Q$ over elements of the $GKBc\gamma$
algebra, it will be useful to write the following representation
of these fields in terms of operators acting on the identity
string field $ |I\rangle=U_{1}^\dag U_{1} |0\rangle$
\begin{eqnarray}
\label{KK} K &\equiv& \frac{1}{2} \hat{\mathcal{L}} U_{1}^\dag
U_{1} |0\rangle,
\\
\label{BB} B &\equiv& \frac{1}{2} \hat{\mathcal{B}} U_{1}^\dag
U_{1} |0\rangle,
\\
\label{GG} G &\equiv& \frac{1}{2} \hat{\mathcal{G}} U_{1}^\dag
U_{1}
|0\rangle, \\
\label{cc} c &\equiv&   U_{1}^\dag U_{1} \tilde c (0)|0\rangle, \\
\label{gg} \gamma &\equiv&   U_{1}^\dag U_{1} \tilde \gamma
(0)|0\rangle.
\end{eqnarray}

The operators $\hat{\mathcal{L}}$, $\hat{\mathcal{B}}$,
$\hat{\mathcal{G}}$, $\tilde c(0)$ and $\tilde \gamma(0)$ are
defined in the sliver frame ($\tilde z$ coordinate)\footnote{To
map a point $z$ in the upper half plane to a point $\tilde z$ in
the sliver frame, we are using the conformal transformation
$\tilde z = \frac{2}{\pi} \arctan z$ \cite{Erler:2009uj}. There is
another convention for the conformal transformation which is given
by $\tilde z = \arctan z$ \cite{Schnabl:2005gv}. In this
convention, instead of the factor $1/2$ in front of the R.H.S. of
equations (\ref{KK})-(\ref{GG}), we should have the factor
$1/\pi$.}, and they are related to the worldsheet energy-momentum
tensor, the $b$ field, the worldsheet supercurrent, the $c$ and
$\gamma$ ghosts fields respectively, for instance
\begin{eqnarray}
\label{Lhat01} \hat{\mathcal{L}} &\equiv& \mathcal{L}_{0}
+\mathcal{L}^{\dag}_0 = \oint \frac{d z}{2 \pi i} (1+z^{2})
(\arctan z+\text{arccot} z) \,
T(z) \, , \\
\label{Bhat01} \hat{\mathcal{B}} &\equiv& \mathcal{B}_{0}
+\mathcal{B}^{\dag}_0 = \oint \frac{d z}{2 \pi i} (1+z^{2})
(\arctan z+\text{arccot} z) \, b(z)
\, , \\
\label{Ghat01} \hat{\mathcal{G}} &\equiv& \mathcal{G}_{1/2}
+\mathcal{G}^{\dag}_{1/2} = \sqrt{\frac{2}{\pi}} \oint \frac{d
z}{2 \pi i} (1+z^{2})^{1/2} (\arctan z+\text{arccot} z) \, G(z) \,
,
\end{eqnarray}
while the operator $U_{1}^\dag U_{1}$ in general is given by
$U^\dag_r U_r = e^{\frac{2-r}{2} \hat{\mathcal{L}}}$, so we have
chosen $r=1$, note that the string field $U_{1}^\dag U_{1}
|0\rangle$ represents to the identity string field. To compute
star products of string fields involving the operators
$\hat{\mathcal{L}}$, $\hat{\mathcal{B}}$ and $\hat{\mathcal{G}}$,
it should be useful to define the operators
\begin{eqnarray}
\label{Lm1} \mathcal{L}_{-1}
&\equiv& \frac{\pi}{2} \oint \frac{d z}{2 \pi i} (1+z^{2})  T(z) = \frac{\pi}{2} (L_{-1}+L_{1}) \, , \\
\label{Bm1} \mathcal{B}_{-1} &\equiv& \frac{\pi}{2} \oint \frac{d
z}{2 \pi i} (1+z^{2})  b(z) = \frac{\pi}{2} (b_{-1}+b_{1})
\, , \\
\label{Gm12} \mathcal{G}_{-1/2} &\equiv& \sqrt{\frac{\pi}{2}}
\oint \frac{d z}{2 \pi i} (1+z^{2})^{1/2} G(z) \, .
\end{eqnarray}

Given two string fields $\phi$ and $\varphi$ belonging to the
$GSO(+)$ or the $GSO(-)$ sector, we can show that
\begin{eqnarray}
\label{StaP1} (\hat{\mathcal{B}}\phi)*\varphi &=&
\hat{\mathcal{B}}(\phi*\varphi) +
(-1)^{\text{gn}(\phi)} \phi * \mathcal{B}_{-1} \varphi \; , \\
\label{StaP2} \phi * (\hat{\mathcal{B}}\varphi)&=&
(-1)^{\text{gn}(\phi)}\hat{\mathcal{B}}(\phi*\varphi)-(-1)^{\text{gn}(\phi)}(\mathcal{B}_{-1}\phi)*\varphi
\; ,  \\
\label{StaP3}
(\hat{\mathcal{B}}\phi)*(\hat{\mathcal{B}}\varphi)&=&-(-1)^{\text{gn}(\phi)}
\hat{\mathcal{B}}\mathcal{B}_{-1}(\phi*\varphi)+
(\mathcal{B}_{-1} \phi)*(\mathcal{B}_{-1} \varphi)\; , \\
\label{StaP4} (\hat{\mathcal{G}}\phi)*\varphi &=&
\hat{\mathcal{G}}(\phi*\varphi) + (-1)^{\text{gn}(\phi)}
 \phi * \mathcal{G}_{-1/2} \varphi \; , \\
\label{StaP5} \phi * (\hat{\mathcal{G}}\varphi)&=&
(-1)^{\text{gn}(\phi)}\hat{\mathcal{G}}(\phi*\varphi)-(-1)^{\text{gn}(\phi)}(\mathcal{G}_{-1/2}\phi)*\varphi
\; , \\
\label{StaP6}
(\hat{\mathcal{G}}\phi)*(\hat{\mathcal{G}}\varphi)&=&-(-1)^{\text{gn}(\phi)}
\hat{\mathcal{G}}\mathcal{G}_{-1/2}(\phi*\varphi)+
(\mathcal{G}_{-1/2} \phi)*(\mathcal{G}_{-1/2} \varphi) \nonumber
\\
&&+(-1)^{\text{gn}(\phi)} 2 \hat{\mathcal{L}} (\phi*\varphi)
+(-1)^{\text{gn}(\phi)}  \phi * \mathcal{L}_{-1} \varphi
\nonumber \\
&&-(-1)^{\text{gn}(\phi)}(\mathcal{L}_{-1}\phi)*\varphi \; ,
\\
\label{StaP7} (\hat{\mathcal{L}}^{n}\phi)*\varphi &=&
\sum_{n'=0}^{n}  {n \choose n'}  \hat{\mathcal{L}}^{n-n'}(\phi*\mathcal{L}_{-1}^{n'}\varphi)\; , \\
\label{StaP8} \phi*(\hat{\mathcal{L}}^{n}\varphi) &=&
\sum_{n'=0}^{n}  {n \choose n'}  (-1)^{n'}
\hat{\mathcal{L}}^{n-n'}((\mathcal{L}_{-1}^{n'}\phi)*\varphi)\; , \\
\label{StaP9}
(\hat{\mathcal{L}}^{m}\phi)*(\hat{\mathcal{L}}^{n}\varphi) &=&
\sum_{m'=0}^{m}\sum_{n'=0}^{n} {m \choose m'} {n \choose n'}
 (-1)^{n'}
\hat{\mathcal{L}}^{m+n-m'-n'}((\mathcal{L}_{-1}^{n'}\phi)*(\mathcal{L}_{-1}^{m'}\varphi))\;
,
\end{eqnarray}
where $\text{gn}(\phi)$ takes into account the Grassmannality of
the string field $\phi$. The above results, containing the
operator $\hat{\mathcal{G}}$, are new and they are an extension of
the result derived in \cite{Schnabl:2005gv}.

Regarding the wedge states with insertions, the star product of
two of them is written in the form
\begin{eqnarray}
\label{StaPWed1} U_{r}^\dag U_{r} \tilde \phi (\tilde x) |0\rangle
* U_{s}^\dag U_{s} \tilde \psi (\tilde y) |0\rangle = U_{t}^\dag
U_{t} \tilde \phi \big(\tilde x + \frac{1}{2}(s-1)\big) \tilde
\psi \big(\tilde y - \frac{1}{2}(r-1)\big) |0\rangle,
\end{eqnarray}
where $t=r+s-1$, and by $\tilde \phi (\tilde x)$ we denote a local
operator $\phi (z)$ expressed in the sliver frame, which in the
special case of primary field with conformal weight $h$ is given
by
\begin{eqnarray}
\label{gf3cor2} \tilde{\phi}(\tilde{z}) =
\big(\frac{dz}{d\tilde{z}}\big)^h \phi (z) =
\big(\frac{\pi}{2}\big)^h \cos^{-2h}\big( \frac{\pi \tilde{z}}{2}
\big) \phi \Big( \tan \big( \frac{\pi \tilde{z}}{2} \big)  \Big).
\end{eqnarray}
Since we are using the conformal transformation $\tilde{z} =
\frac{2}{\pi} \arctan z$ which is a bit different from the one
used in Schnabl's original paper $\tilde{z} = \arctan z$
\cite{Schnabl:2005gv}, we have a factor $1/2$ in the R.H.S. of
equation (\ref{StaPWed1}) instead of the factor $\pi/4$ which is
present in the reference \cite{Schnabl:2005gv}.

It will be useful to know the action of the BRST,
$\mathcal{L}_{-1}$, $\mathcal{B}_{-1}$ and $\mathcal{G}_{-1/2}$
operators on the star product of two string fields
\begin{eqnarray}
\label{Qact1} Q (\phi*\varphi) &=& (Q \phi)*\varphi +
(-1)^{\text{gn}(\phi)}\phi *(Q
\varphi), \\
\label{Qact2} \mathcal{L}_{-1} (\phi*\varphi) &=&
(\mathcal{L}_{-1} \phi)*\varphi + \phi *(\mathcal{L}_{-1}
\varphi),
\\
\label{Qact3} \mathcal{B}_{-1} (\phi*\varphi) &=&
(\mathcal{B}_{-1} \phi)*\varphi +
(-1)^{\text{gn}(\phi)}\phi *(\mathcal{B}_{-1} \varphi), \\
\label{Qact4} \mathcal{G}_{-1/2} (\phi*\varphi) &=&
(\mathcal{G}_{-1/2} \phi)*\varphi + (-1)^{\text{gn}(\phi)}\phi
*(\mathcal{G}_{-1/2} \varphi) \, .
\end{eqnarray}

Let us derive the algebra associated to the set of operators
defined by equations (\ref{KK})-(\ref{gg}). As a pedagogical
illustration, we explicitly compute the product $G^2$
\begin{eqnarray}
\label{GG1} G^2 \equiv G*G = \frac{1}{2} \hat{\mathcal{G}}
U_{1}^\dag U_{1} |0\rangle * \frac{1}{2} \hat{\mathcal{G}}
U_{1}^\dag U_{1} |0\rangle = \frac{1}{4} \hat{\mathcal{G}}
U_{1}^\dag U_{1} |0\rangle *  \hat{\mathcal{G}} U_{1}^\dag U_{1}
|0\rangle,
\end{eqnarray}
using equation (\ref{StaP6}) and the commutators
$[\mathcal{G}_{-1/2},\hat{\mathcal{L}}]=0$,
$[\mathcal{L}_{-1},\hat{\mathcal{L}}]=0$, we obtain
\begin{eqnarray}
\label{GG2} G^2 = \frac{2}{4} \hat{\mathcal{L}} \Big( U_{1}^\dag
U_{1} |0\rangle * U_{1}^\dag U_{1} |0\rangle \Big) = \frac{1}{2}
\hat{\mathcal{L}} U_{1}^\dag U_{1} |0\rangle,
\end{eqnarray}
therefore we have that $G^2=K$.

Following the same steps, using equations
(\ref{StaP1})-(\ref{StaP9}), the commutator relation
$[\mathcal{G}_{-1/2},\tilde \gamma(0)]= - \frac{1}{2}
\partial \tilde c(0)$ and the anti-commutator
$\{\mathcal{G}_{-1/2},\tilde c(0)\}= - 2 \tilde \gamma(0)$, we can
show that
\begin{align}
\label{GG3} \{G,G\} &= 2 K, \;\;\; [K,B]=0, \;\;\; [K,G]=0, \;\;\; \{B,G\} = 0,\\
\label{BG1}  \partial c &= [K,c], \;\;\; \partial \gamma = [K,\gamma],  \;\;\; B^2 = 0, \;\;\; c^2 = 0, \\
\label{B2} \{G,c\} & =- 2 \gamma, \;\;\; [G,\gamma]= -\frac{1}{2}
\partial c \, ,
\end{align}
where the expressions $\partial c$ and $\partial \gamma$ have been
defined as $\partial \phi \equiv U_{1}^\dag U_{1}
\partial \tilde \phi (0) |0\rangle $.

The action of the BRST operator $Q$ on the basic string fields
$K$, $G$, $B$, $c$ and $\gamma$ is given by
\begin{align}
\label{QK} Q K &=0, \;\;\; Q G =0, \;\;\; Q B =K, \\
\label{Qc} Q c &= cKc-\gamma^2, \\
\label{Qg} Q \gamma &= c \partial \gamma - \frac{1}{2} \gamma
\partial c \, .
\end{align}

Now we are in position to study and present the construction of a
one-parameter family of solutions.

\section{One-parameter family of solutions from an identity based solution}
It is known that a solution to the equation of motion
(\ref{EqmC1}) is given by the following simplest identity based
solution
\cite{Arroyo:2010fq,Arefeva:2010yd,AldoArroyo:2012if,Arroyo:2013pha}
\begin{align}
\label{Iden1}  \widehat{\Phi}_I = \Big(  (c+B \gamma^2)(1-K) \Big)
\otimes \sigma_3.
\end{align}
Using this identity based solution (\ref{Iden1}), we will show that it is possible to construct a one-parameter family of solutions
$\widehat{\Phi}_{\lambda}$ which depending on the value of the parameter $\lambda$ will describe three distinct gauge orbits corresponding to the
perturbative vacuum, the half brane and the tachyon vacuum solution.

Let us write the explicit form of the aforementioned gauge
transformation
\begin{eqnarray}
\label{solphi1} \widehat{\Phi}_{\lambda}=
\widehat{U}_{\lambda}(\widehat{Q}+\widehat{\Psi}_I)\widehat{U}^{-1}_{\lambda},
\end{eqnarray}
$\widehat{U}_{\lambda}$ is a ghost number zero string field given
by\footnote{We would like to bring few motivational words
explaining the choice (\ref{gaugeU1}). As in the bosonic case
\cite{Arroyo:2010fq}, for superstring field theory we can also
construct a gauge transformation which relates the identity based
solution (\ref{Iden1}) with the half brane solution
\cite{Erler:2010pr}. The gauge transformation which does this job
precisely corresponds to a $\widehat{U}$ given by
\begin{eqnarray}
\label{gaugefootnote} \widehat{U} = \Big( 1+cB[K-1] \Big) \otimes \mathbb{I} + i cBG \otimes \sigma_1 .
\end{eqnarray}
Applying a supersymmetric analog of the Zeze map
\cite{Zeze:2010sr}, we consider a slight modification of
(\ref{gaugefootnote}) in which a real parameter $\lambda$ is
inserted in the $cBK$ and $cBG$ pieces in the gauge transformation
such that for $\lambda=0$ and $\lambda=1$, we recover the
perturbative and tachyon vacua respectively.}
\begin{eqnarray}
\label{gaugeU1} \widehat{U}_{\lambda} &=& \Big(
1+cB[K+(\lambda-1)(2\lambda+1)] \Big)
\otimes \mathbb{I} + 4i\lambda(1-\lambda)cBG \otimes \sigma_1 , \\
\label{gaugeU2} \widehat{U}^{-1}_{\lambda} &=& \Big(
1-cB\frac{K-1+f(K,\lambda)}{K} \Big) \otimes \mathbb{I} -
c\frac{\widetilde{f}(K,\lambda)}{K}BG \otimes \sigma_1 ,
\end{eqnarray}
where $f(K,\lambda)$ and $\widetilde{f}(K,\lambda)$ are the
following functions
\begin{eqnarray}
\label{gaugeF1} f(K,\lambda) &=&  \frac{\lambda^2 (1-2 \lambda
)^2+\left(16 \lambda^3-32 \lambda^2+18 \lambda -1\right) \lambda
\, K }{\lambda ^2(1-2
   \lambda )^2 +2 \lambda  \left(8 \lambda ^3-16 \lambda ^2+10 \lambda -1\right) K+K^2} \; , \\
\label{gaugeF2} \widetilde{f}(K,\lambda) &=&   \frac{4 i
(1-\lambda) \lambda \, K}{\lambda ^2(1-2
   \lambda )^2 +2 \lambda  \left(8 \lambda ^3-16 \lambda ^2+10 \lambda -1\right) K+K^2} \;
   .
\end{eqnarray}

Then, the one-parameter family of solutions is obtained by
performing the above gauge transformation over the identity based
solution (\ref{Iden1})
\begin{eqnarray}
\widehat{\Phi}_{\lambda} &=&
\widehat{U}_{\lambda}\widehat{Q}\widehat{U}^{-1}_{\lambda} +
\widehat{U}_{\lambda} \Big(  (c+B \gamma^2)(1-K)
\otimes \sigma_3 \Big) \widehat{U}^{-1}_{\lambda} \nonumber \\
\label{solphi2} &=& \Phi_{1,\lambda} \otimes \sigma_3 +
\Phi_{2,\lambda} \otimes i \sigma_2,
\end{eqnarray}
where the string fields $\Phi_{1,\lambda}$ and $\Phi_{2,\lambda}$
are given by
\begin{eqnarray}
\label{solphi3} \Phi_{1,\lambda} &=& Q(Bc)f(K,\lambda) + \lambda(2\lambda-1)c f(K,\lambda) + 4i\lambda(1-\lambda) c GB c G \widetilde{f}(K,\lambda) ,\\
\label{solphi4} \Phi_{2,\lambda} &=& Q(Bc) G
\widetilde{f}(K,\lambda) + \lambda(2\lambda-1)c G
\widetilde{f}(K,\lambda) + 4i \lambda(1-\lambda) c G B c
f(K,\lambda).
\end{eqnarray}
A check of the equation of motion for the above solution is
straightforward.

At this point we can ask about the interval where the parameter $\lambda$ should belong,
the answer to this question will be studied later, for the time
being, let us analyze the solution for particular values of this parameter.

For the value of the parameter $\lambda=0$, we identically obtain
$\widehat{\Phi}_{\lambda=0}=0$ and thus this case corresponds to
the perturbative vacuum.

For the value $\lambda = 1$, we see that $
\widetilde{f}(K,\lambda=1)=0$ and $f(K,\lambda=1)=1/(1+K)$,
therefore we obtain
\begin{eqnarray}
\label{sollam1} \widehat{\Phi}_{\lambda=1}= \big[ Q(Bc) + c
\big]\frac{1}{1+K} \otimes \sigma_3.
\end{eqnarray}
This solution precisely represents the tachyon vacuum solution. The energy of this solution (\ref{sollam1}) has been evaluated in references
\cite{Gorbachev:2010zz,Arroyo:2010fq} given a result in agreement with Sen's first conjecture.

For the value $\lambda = 1/2$, we get $
\widetilde{f}(K,\lambda=1/2)=i/(1+K)$ and
$f(K,\lambda=1/2)=1/(1+K)$, so in this case the solution can be
written as
\begin{eqnarray}
\label{sollam2} \widehat{\Phi}_{\lambda=1/2}= \Big[ Q(Bc) - c GB c
G \Big]\frac{1}{1+K} \otimes \sigma_3 + \Big[ i Q(Bc)G +i c GB c
\Big]\frac{1}{1+K} \otimes i \sigma_2  .
\end{eqnarray}
This solution has been studied in reference \cite{Erler:2010pr}
and since the evaluation of its energy brings a result which is
half of the value of the tachyon vacuum energy, the solution
(\ref{sollam2}) has been called as the half brane solution.

Note that to recognize the kind of solution we have, we must calculate the energy associated to the solution. For any solution of the form $\widehat{\Phi}
= \Phi_1 \otimes \sigma_3 + \Phi_2 \otimes i \sigma_2$, employing equation (\ref{action2}), we can write the normalized value of the energy $E$ as follows
\begin{eqnarray}
\label{NorV1} E(\widehat{\Phi}) \equiv -2 \pi^2 g^2 S =
\frac{\pi^2}{3} \Big[\langle Y_{-2} \Phi_1 Q \Phi_1 \rangle +
\langle Y_{-2} \Phi_2 Q \Phi_2 \rangle \Big].
\end{eqnarray}
To evaluate the energy (\ref{NorV1}) for the solution
(\ref{solphi2}) with a generic value of the parameter $\lambda$,
we will require to define and study correlation functions
involving elements of the $GKBc\gamma$ algebra. In the next
section, we are going to consider correlation functions including
the $G$ field and as a pedagogical application of these
correlators, we will show the computation of the energy for the
half brane solution.

\section{Correlation functions and the half brane energy}
To compute the energy for solutions constructed out of elements of the $GKBc\gamma$ algebra, it will be useful to know correlation functions defined on a
semi-infinite cylinder of circumference $l$ denoted by $C_{l}$.

A point $z$ on the upper half-plane can be mapped to a point $\tilde z \in C_{l}$, which has the property that $\tilde z \simeq \tilde z + l$, through the
conformal transformation
\begin{eqnarray}
\label{gf2cor1P} \tilde{z} = \frac{l}{\pi} \arctan z,
\end{eqnarray}
The expression for the conformal transformation of primary fields with conformal weight $h$ is given by
\begin{eqnarray}
\label{gf3cor2P} \tilde{\phi}(\tilde{z}) =
\big(\frac{dz}{d\tilde{z}}\big)^h \phi (z) =
\big(\frac{\pi}{l}\big)^h \cos^{-2h}\big( \frac{\pi \tilde{z}}{l}
\big) \phi \Big( \tan \big( \frac{\pi \tilde{z}}{l} \big)  \Big).
\end{eqnarray}
Using (\ref{gf2cor1P}) and (\ref{gf3cor2P}), we can derive the following correlation function involving the $b(z)$, $c(z)$ and $\gamma(z)$ ghost fields
\begin{align}
\label{gf8cor3P1} \langle Y_{-2} c(\tilde{x})
\gamma(\tilde{y})\gamma(\tilde{z}) \rangle_{C_l} &= \frac{l^2}{2
\pi^2} \cos \Big(  \frac{\pi (\tilde{y}-\tilde{z})}{l} \Big), \\
\label{gf8cor3P2} \langle Y_{-2} b(\tilde{v}) c(\tilde{w})
c(\tilde{x}) \gamma(\tilde{y})\gamma(\tilde{z}) \rangle_{C_l} &=
\frac{l \csc \left(\frac{\pi (\tilde{v}-\tilde{w})}{l}\right) \csc
\left(\frac{\pi
   (\tilde{v}-\tilde{x})}{l}\right) \sin \left(\frac{\pi  (\tilde{w}-\tilde{x})}{l}\right) \cos \left(\frac{\pi
   (\tilde{y}-\tilde{z})}{l}\right)}{2 \pi } .
\end{align}

Using (\ref{gf8cor3P2}), let us compute the correlator $\langle
Y_{-2} B c(\tilde{w}) c(\tilde{x})
\gamma(\tilde{y})\gamma(\tilde{z}) \rangle_{C_l}$. Since the $B$
field can be defined as a line integral insertion of the $b(z)$
ghost field inside correlation functions on the cylinder
\cite{Okawa:2006vm}, we can write
\begin{align}
\label{way3} \langle Y_{-2} B  c(\tilde{w}) c(\tilde{x})
\gamma(\tilde{y})\gamma(\tilde{z}) \rangle_{C_l} = \langle Y_{-2}
\int_{-i \infty}^{i\infty} \frac{d \tilde{v}}{2 \pi i}
b(\tilde{v}) c(\tilde{w}) c(\tilde{x})
\gamma(\tilde{y})\gamma(\tilde{z}) \rangle_{C_l}.
\end{align}
Plugging (\ref{gf8cor3P2}) into the R.H.S. of equation (\ref{way3}) and employing the integral
\begin{eqnarray}
\label{InA1} \int_{-i\infty}^{i \infty} d \tilde v \,\csc
\left(\frac{\pi (\tilde{v}-\tilde{w})}{l}\right) \csc
\left(\frac{\pi
   (\tilde{v}-\tilde{x})}{l}\right)=2i(\tilde{w}-\tilde{x}) \csc
\left(\frac{\pi (\tilde{w}-\tilde{x})}{l}\right),
\end{eqnarray}
we obtain
\begin{align}
\label{CBccgg1} \langle Y_{-2} B  c(\tilde{w}) c(\tilde{x})
\gamma(\tilde{y})\gamma(\tilde{z}) \rangle_{C_l} = \frac{l}{2
\pi^2} (\tilde{w}-\tilde{x}) \cos \Big(  \frac{\pi
(\tilde{y}-\tilde{z})}{l} \Big).
\end{align}

In the same way, by writing the $G$ field as a line integral insertion of the worldsheet supercurrent $G(z)$ inside correlation functions on the cylinder,
we can derive the following correlators
\begin{eqnarray}
\label{Gccg1} \langle Y_{-2} G c(\tilde x)c(\tilde y)
\gamma(\tilde z) \rangle_{C_l} = \frac{l^2}{2 \pi ^2} \Big[ \cos
\left(\frac{\pi
   \left(\tilde y-\tilde z \right)}{l}\right) - \cos
   \left(\frac{\pi  \left(\tilde x-\tilde z \right)}{l}\right)
   \Big],
\end{eqnarray}
\begin{align}
\label{GBcccg1} \langle Y_{-2} G B c(\tilde w) c(\tilde x)
c(\tilde y) \gamma(\tilde z) \rangle_{C_l} =
\;\;\;\;\;\;\;\;\;\;\;\;
\;\;\;\;\;\;\;\;\;\;\;\;\;\;\;\;\;\;\;\;\;\;\;\;\;\;\;\;\;\;\;\;\;\;\;\;\;\;
\;\;\;\;\;\;\;\;\;\;\;\;\;\;\;\;\;\;\;\;\;\;\;\;\;\;\;\;\;\;\;\;\; \nonumber \\
=\frac{l \left((\tilde{x}-\tilde{y}) \cos \left(\frac{\pi
   (\tilde{w}-\tilde{z})}{l}\right)+(\tilde{y}-\tilde{w}) \cos \left(\frac{\pi
   (\tilde{x}-\tilde{z})}{l}\right)+(\tilde{w}-\tilde{x}) \cos \left(\frac{\pi
   (\tilde{y}-\tilde{z})}{l}\right)\right)}{2 \pi ^2} \, ,
\end{align}
\begin{align}
\label{GBcggg1} \langle Y_{-2} G B c(\tilde w) \gamma(\tilde
x)\gamma (\tilde y) \gamma(\tilde z) \rangle_{C_l} = \frac{l
\left(\cos \left(\frac{\pi (\tilde{x}-\tilde{y})}{l}\right)+\cos
   \left(\frac{\pi  (\tilde{x}-\tilde{z})}{l}\right)+\cos \left(\frac{\pi
   (\tilde{y}-\tilde{z})}{l}\right)\right)}{8 \pi ^2}.
\end{align}

With the aid of these correlation functions, we are ready to
evaluate the energy associated to the half brane solution. Using
equation (\ref{NorV1}) for the particular case of the solution
(\ref{sollam2}), and noting that the BRST exact terms do not
contribute to the evaluation of the energy, we obtain
\begin{align}
\label{NorV1H1} E(\widehat{\Phi}_{\lambda=1/2}) = \frac{\pi^2}{3}
\Big[\langle\langle cGBcG \frac{1}{1+K} Q(cGBc) G \frac{1}{1+K}
\rangle \rangle - \langle \langle cGBc \frac{1}{1+K} Q(cGBc)
\frac{1}{1+K} \rangle \rangle \Big],
\end{align}
where the notation $\langle \langle \cdots \rangle \rangle$ means
that $\langle \langle \cdots \rangle \rangle \equiv \langle Y_{-2}
\cdots \rangle$. Employing equations (\ref{GG3})-(\ref{Qg}), after
a lengthy algebraic manipulations, from equation (\ref{NorV1H1})
we arrive to
\begin{align}
\label{NorV1H2} E(\widehat{\Phi}_{\lambda=1/2}) = \frac{\pi^2}{3}
\Big[\langle\langle Kc\frac{1}{1+K}\gamma^2 \frac{1}{1+K}\rangle
\rangle + 3 \langle\langle  KcK\frac{1}{1+K}\gamma^2 \frac{1}{1+K}
\rangle \rangle -\frac{2}{3} \langle\langle
GcK^2\frac{1}{1+K}c\gamma \frac{1}{1+K}\rangle \rangle  \nonumber
\\ + \langle\langle G\gamma \frac{1}{1+K}cKc\frac{1}{1+K}\rangle
\rangle -5 \langle\langle BcKc\frac{1}{1+K}\gamma^2
\frac{1}{1+K}\rangle \rangle-4 \langle\langle Bc\gamma
K\frac{1}{1+K}c\gamma \frac{1}{1+K}\rangle \rangle \nonumber \\
+2 \langle\langle B c \gamma K^2 \frac{1}{1+K}c\gamma
\frac{1}{1+K}\rangle \rangle + 4 \langle\langle G B c\gamma
\frac{1}{1+K}\gamma^2 \frac{1}{1+K}\rangle \rangle-6
\langle\langle B c \gamma K \frac{1}{1+K} c \gamma
K\frac{1}{1+K}\rangle \rangle \nonumber \\
+ 4 \langle\langle GBcK\frac{1}{1+K}\gamma^3 \frac{1}{1+K}\rangle
\rangle-3 \langle\langle GBc\frac{1}{1+K}cKc\gamma
   \frac{1}{1+K}\rangle \rangle-3 \langle\langle GBcK\frac{1}{1+K}cKc\gamma \frac{1}{1+K}\rangle
   \rangle \Big]
\end{align}
All the above correlators can be computed using equations
(\ref{gf8cor3P1}) and (\ref{CBccgg1})-(\ref{GBcggg1}), for
instance, let us explicitly compute the correlator $\langle\langle
GBcK\frac{1}{1+K}cKc\gamma \frac{1}{1+K}\rangle \rangle $
\begin{align}
\label{NorV1H2A1} \langle\langle GBcK\frac{1}{1+K}cKc\gamma
\frac{1}{1+K}\rangle \rangle = \int_{0}^{\infty} dt_1
dt_2\,e^{-t_1-t_2} \partial_{s_1} \partial_{s_2}
\Big[\langle\langle GBc\Omega^{s_1+t_1}c\Omega^{s_2}c\gamma
\Omega^{t_2} \rangle \rangle \Big] \Big{|}_{s_1=s_2=0},
\end{align}
where we have used the fact that $\Omega^t = e^{-tK}$. The
correlator $\langle\langle GBc\Omega^{s_1+t_1}c\Omega^{s_2}c\gamma
\Omega^{t_2} \rangle \rangle$ is given by
\begin{align}
\label{NorV1H2A2} \langle\langle
GBc\Omega^{s_1+t_1}c\Omega^{s_2}c\gamma \Omega^{t_2} \rangle
\rangle = \langle Y_{-2}
GBc(s_1+s_2+t_1+t_2)c(s_2+t_2)c(t_2)\gamma(t_2)
\rangle_{C_{s_1+s_2+t_1+t_2}} .
\end{align}
The R.H.S. of equation (\ref{NorV1H2A2}) can be evaluated using
equation (\ref{GBcccg1}), so that we obtain the result
\begin{align}
\label{NorV1H2A3} \partial_{s_1} \partial_{s_2}
\Big[\langle\langle GBc\Omega^{s_1+t_1}c\Omega^{s_2}c\gamma
\Omega^{t_2} \rangle \rangle \Big] \Big{|}_{s_1=s_2=0} =
\;\;\;\;\;\;\;\;\;\;\;\;\;\;\;\;\;\;\;\;\;\;\;\;\;\;\;
\;\;\;\;\;\;\;\;\;\;\;\;\;\;\;\;\;\;\;\;\;\;\;\;\;\;\; \nonumber
\\ = \frac{t_1 \left(\cos \left(\frac{\pi
t_1}{t_1+t_2}\right)-1\right)+t_2 \left(-\pi  \sin \left(\frac{\pi
   t_1}{t_1+t_2}\right)+\cos \left(\frac{\pi  t_1}{t_1+t_2}\right)-1\right)}{2 \pi ^2
   \left(t_1+t_2\right)} \, .
\end{align}
Performing the change of variables $t_1 \rightarrow  u v$, $t_2
\rightarrow  u - u v$, $\int_{0}^{\infty} dt_1 dt_2 \rightarrow
\int_{0}^{\infty} du \int_{0}^{1} dv \, u  $, and using the result
(\ref{NorV1H2A3}), from equation (\ref{NorV1H2A1}), we get
\begin{align}
\label{NorV1H2A4} \langle\langle GBcK\frac{1}{1+K}cKc\gamma
\frac{1}{1+K}\rangle \rangle &= \int_{0}^{\infty} du \int_{0}^{1}
dv \, \frac{e^{-u} u (\pi  (v-1) \sin (\pi  v)+\cos (\pi  v)-1)}{2
\pi ^2} \nonumber \\
&= -\frac{1}{\pi ^2} \, .
\end{align}
Performing similar computations for the rest of terms appearing on
the R.H.S. of equation (\ref{NorV1H2}) and adding the results up,
the energy turns out to be
\begin{align}
\label{NorV1H2A5} E(\widehat{\Phi}_{\lambda=1/2}) =
\frac{\pi^2}{3} \Big[ - \frac{3}{2 \pi^2} \Big] = - \frac{1}{2},
\end{align}
this is precisely $1 / 2$ times the normalized value of the
tachyon vacuum energy which has the value
$E(\widehat{\Phi}_{\lambda=1}) = -1$.

Let us summarize the results for the normalized value of the energy (\ref{NorV1}) which has been obtained for the particular values of the parameter
$\lambda=\{0,1/2,1\}$
\begin{align}
\label{NorV1H2A6} E(\widehat{\Phi}_{\lambda}) = \begin{cases}
      0, & \lambda = 0 \;,\;\; \text{Perturbative Vacuum Solution,} \\
      -1/2, & \lambda = 1/2 \;,\;\; \text{Half Brane Solution,} \\
      -1, & \lambda = 1 \;,\;\; \text{Tachyon Vacuum Solution.}
   \end{cases}
\end{align}

Finally, we would like to evaluate the energy
$E(\widehat{\Phi}_{\lambda})$ for a generic value of the parameter
$\lambda$. This computation will be performed in the next section.

\section{Energy of the one-parameter family of solutions}
In order to evaluate the energy associated to the one-parameter
family of solutions $\widehat{\Phi}_{\lambda}$ for a generic value
of the parameter $\lambda$, it will be useful to express the
functions (\ref{gaugeF1}) and (\ref{gaugeF2}) as superpositions of
wedge states $\Omega^t = e^{-tK} $, to this end, let us start by
rewriting the solution (\ref{solphi2}) as follows
\begin{eqnarray}
\label{solphi2xd} \widehat{\Phi}_{\lambda}= \Phi_{1,\lambda}
\otimes \sigma_3 + \Phi_{2,\lambda} \otimes i \sigma_2,
\end{eqnarray}
where the $GSO(\pm)$ components $\Phi_{1,\lambda}$ and
$\Phi_{2,\lambda}$ are given by
\begin{eqnarray}
\label{solphi3xd} \Phi_{1,\lambda} &=& Q(Bc)f(K,\lambda) + p c f(K,\lambda) + q c GB c G \widetilde{f}(K,\lambda) ,\\
\label{solphi4xd} \Phi_{2,\lambda} &=& Q(Bc) G
\widetilde{f}(K,\lambda) +p c G \widetilde{f}(K,\lambda) + q c G B
c f(K,\lambda),
\end{eqnarray}
and
\begin{eqnarray}
\label{gaugeF1xd} f(K,\lambda) &=&  \frac{p^2 +w K }{(K-r_1)(K-r_2)} \; , \\
\label{gaugeF2xd} \widetilde{f}(K,\lambda) &=&   \frac{q
K}{(K-r_1)(K-r_2)} \;
   .
\end{eqnarray}
The set of parameters $p$, $q$, $w$, $r_1$ and $r_2$ have been
defined as
\begin{align}
\label{defpqw} p = \lambda(2\lambda-1), \;\;\;\;\;\;\;\; q = 4i
\lambda(1-\lambda), \;\;\;\;\;\;\;\; w =  \lambda (16 \lambda^3-32
\lambda^2+18 \lambda -1), \\
\label{defr1} r_1 = -8 \lambda ^4+16 \lambda ^3-10 \lambda
^2+\lambda-4 \sqrt{4 \lambda ^8-16 \lambda ^7+26
   \lambda ^6-21 \lambda ^5+8 \lambda ^4-\lambda ^3}, \\
\label{defr2} r_2 = -8 \lambda ^4+16 \lambda ^3-10 \lambda
^2+\lambda+4 \sqrt{4 \lambda ^8-16 \lambda ^7+26
   \lambda ^6-21 \lambda ^5+8 \lambda ^4-\lambda ^3}.
\end{align}

Using partial fraction decomposition, the functions defined by
equations (\ref{gaugeF1xd}) and (\ref{gaugeF2xd}) can be expressed
as
\begin{eqnarray}
\label{gaugeF1xd22} f(K,\lambda) &=&  \frac{\alpha_1}{K-r_1} + \frac{\beta_1}{K-r_2} \; , \\
\label{gaugeF2xd22} \widetilde{f}(K,\lambda) &=&
\frac{\alpha_2}{K-r_1} + \frac{\beta_2}{K-r_2} \;
   ,
\end{eqnarray}
where the parameters $\alpha_1$, $\alpha_2$, $\beta_1$ and
$\beta_2$ are given by
\begin{align}
\label{alphabeta1} \alpha_1 &= \frac{p^2+r_1 w}{r_1-r_2} \, , \;\;\;\; \beta_1 = -\frac{p^2+r_2 w}{r_1-r_2} \, , \\
\label{alphabeta2} \alpha_2 &= \frac{q r_1}{r_1-r_2} \, ,
\;\;\;\;\;\;\; \beta_2 =-\frac{q r_2}{r_1-r_2} \, .
\end{align}

The way how we have written the functions (\ref{gaugeF1xd22}) and
(\ref{gaugeF2xd22}) allow us to represent these functions as the
following integrals
\begin{eqnarray}
\label{gaugeF1xd33} f(K,\lambda) &=&  \int_{0}^{\infty} dt \big[\alpha_1 e^{r_1 t}+ \beta_1 e^{r_2 t}\big] \Omega^{t} \; , \\
\label{gaugeF2xd33} \widetilde{f}(K,\lambda) &=& \int_{0}^{\infty}
dt \big[\alpha_2 e^{r_1 t}+ \beta_2 e^{r_2 t}\big] \Omega^{t} \; .
\end{eqnarray}
This integral representation constitutes a superposition of wedge states $\Omega^t = e^{-tK}$ \cite{Schnabl:2010tb}.

In order for these integrals (\ref{gaugeF1xd33}) and
(\ref{gaugeF2xd33}) to provide convergent results, we should
require
\begin{eqnarray}
\label{condilam1} \big(\Re r_1 < 0\big) \wedge \big(\Re r_2 <
0\big).
\end{eqnarray}
Using equations (\ref{defr1}) and (\ref{defr2}), from this
inequality (\ref{condilam1}) we obtain the following conditions
for the parameter $\lambda$
\begin{eqnarray}
\label{condilam2} \big(\lambda < 0\big) \vee \big(\kappa \leq
\lambda <\frac{1}{2}\big) \vee \big(\lambda >\frac{1}{2}\big),
\end{eqnarray}
where $\kappa$ is a numerical constant defined as
\begin{eqnarray}
\label{kappa1} \kappa = \frac{2}{3}-\frac{1}{6}
\left(\frac{25}{2}+\frac{3}{2} \sqrt{69}\right)^{1/3}-\frac{1}{6}
\left(\frac{25}{2}-\frac{3}{2} \sqrt{69}\right)^{1/3} \approx
0.122561
\end{eqnarray}

It is interesting to note that the region (\ref{condilam2}) does not contain the points $\lambda=0$ and $\lambda = 1/2$ which corresponds to the
perturbative vacuum and the half brane solution respectively. Physically this means that the various values: $\lambda=0$, $\lambda=1/2$ and the ones
defined by the region (\ref{condilam2}) formally correspond to distinct gauge orbits within the formal solution (\ref{solphi1}).

Now, we are going to evaluate the energy $ E(\widehat{\Phi}_{\lambda})$ associated to a parameter $\lambda$ belonging to the region (\ref{condilam2}). We
might anticipate the result using the following argument. Due to the fact that the energy is a gauge invariant quantity, and since $\lambda$ belonging to
the region (\ref{condilam2}) corresponds to an specific gauge orbit, to compute the energy, we can choose a particular value for the parameter $\lambda$
contained in this region, for instance $\lambda=1$ which we know corresponds to the tachyon vacuum solution, therefore we should obtain the following
result for the energy
\begin{eqnarray}
 \label{Earg1} E(\widehat{\Phi}_{\lambda}) = -1 \, , \;\;\; \text{for} \;\; \big(\lambda < 0\big) \vee \big(\kappa\leq
\lambda <\frac{1}{2}\big) \vee \big(\lambda >\frac{1}{2}\big).
\end{eqnarray}
Employing the solution (\ref{solphi2xd}) together with the
integral representation of the functions $f$ and $\widetilde{f}$
given by (\ref{gaugeF1xd33}) and (\ref{gaugeF2xd33}), we would
like to check the validity of the above result.

Using equation (\ref{NorV1}) for the case of the solution
(\ref{solphi2xd}), and noting that the BRST exact terms do not
contribute to the evaluation of the energy, we obtain
\begin{align}
\label{NorV1H1Sollam1} E(\widehat{\Phi}_{\lambda}) =
\frac{\pi^2}{3} \Big[p^2 \langle\langle c f Q(c)f  \rangle \rangle
+ q^2 \langle\langle c GB c G \widetilde{f} Q(c GB c) G
\widetilde{f} \rangle \rangle + 2 p q \langle\langle c GB c G
\widetilde{f} Q(c)f \rangle \rangle \nonumber \\
+ p^2 \langle\langle c G \widetilde{f} Q(c)G \widetilde{f} \rangle
\rangle + q^2 \langle\langle c GB c  f Q(c GB c) f \rangle \rangle
+ 2 p q \langle\langle c GB c f Q(c)G \widetilde{f} \rangle
\rangle \Big].
\end{align}
Employing the identities (\ref{GG3})-(\ref{Qg}), the correlation functions (\ref{gf8cor3P1}), (\ref{CBccgg1})-(\ref{GBcggg1}), the integrals
(\ref{gaugeF1xd33}) and (\ref{gaugeF2xd33}), we can evaluate all the correlation functions which will appear from the R.H.S. of (\ref{NorV1H1Sollam1}). For
instance, let us compute $\langle\langle c f Q(c)f  \rangle \rangle$
\begin{align}
\label{AcorrA1} \langle\langle c f Q(c)f  \rangle \rangle =
-\langle Y_{-2} c f(K,\lambda) \gamma^2 f(K,\lambda)  \rangle = -
\int_{0}^{\infty} dt_1 dt_2 \, h(t_1) h(t_2) \langle Y_{-2} c
\Omega^{t_1} \gamma^2 \Omega^{t_2}  \rangle ,
\end{align}
where we have defined $h(t) = \alpha_1 e^{r_1 t}+ \beta_1 e^{r_2
t}$. Using the correlation function (\ref{gf8cor3P1}), we can
derive the correlator
\begin{eqnarray}
\label{AcorrA2} \langle Y_{-2} c \Omega^{t_1} \gamma^2
\Omega^{t_2} \rangle = \frac{(t_1+t_2)^2}{2 \pi^2}.
\end{eqnarray}
Plugging (\ref{AcorrA2}) into equation (\ref{AcorrA1}) and
performing the change of variables $t_1 \rightarrow  u v$, $t_2
\rightarrow  u - u v$, $\int_{0}^{\infty} dt_1 dt_2 \rightarrow
\int_{0}^{\infty} du \int_{0}^{1} dv \, u  $, we get
\begin{align}
\label{AcorrA3} \langle\langle c f Q(c)f  \rangle \rangle  &= -
\int_{0}^{\infty} du \int_{0}^{1} dv \, \frac{u^3 \big(\alpha _1
e^{r_1 u v}+\beta _1 e^{r_2 u v}\big) \big(\alpha _1 e^{r_1 (u-u
v)}+\beta _1
   e^{r_2 (u-u v)}\big)}{2 \pi^2}   \\
&= -\frac{2 \alpha _1 \beta _1 r_1 r_2 (r_1^2+r_2 r_1+r_2^2) +3
\alpha _1^2 r_2^4+3 \beta _1^2
   r_1^4}{\pi ^2 r_1^4 r_2^4} \nonumber \\
\label{AcorrA3Aux} &= \frac{3 - 38 \lambda + 64 \lambda^2 - 32
\lambda^3}{\pi^2 \lambda^2 (2 \lambda-1)^3} \, .
\end{align}
The integral (\ref{AcorrA3}) exists only when $\Re \, r_{1,2} < 0$, and for such $r_1,r_2$, this integral has the value shown in equation
(\ref{AcorrA3Aux}). Note that we have a singularity at $\lambda =0$ and $\lambda = 1/2$, while in the case where $\lambda$ belongs to the region
$(0,\kappa)$, the expression (\ref{AcorrA3Aux}) is clearly well-defined. Therefore aside from these two singular points, it seems that the result of the
integral does not differentiate between different regions of $\lambda$. We wonder if the same phenomenon can happen for the remaining integrals coming from
the rest of terms on the R.H.S. of equation (\ref{NorV1H1Sollam1}).

It turns out that the expressions for the remaining integrals will not be as simple as the one shown in (\ref{AcorrA3}). For instance, from the second term
on the R.H.S. of equation (\ref{NorV1H1Sollam1}), after performing algebraic manipulations, we obtain a lot of terms and just as an illustration, let us
show one of them
\begin{align}
\label{AcorrExtra1} \mathcal{I}(\lambda) \equiv \langle Y_{-2} B K
c \widetilde{f}(K,\lambda) \gamma K \widetilde{f}(K,\lambda) c
\gamma \rangle.
\end{align}
Using the integral representation (\ref{gaugeF2xd33}), and
defining the function $g(t) = \alpha_2 e^{r_1 t}+ \beta_2 e^{r_2
t}$, we can write equation (\ref{AcorrExtra1}) as follows
\begin{align}
\label{AcorrExtra2}  \mathcal{I}(\lambda) = \int_{0}^{\infty} dt_1
dt_2 \, g(t_1) g(t_2) \langle Y_{-2} B K c \Omega^{t_1} \gamma K
\Omega^{t_2} c \gamma \rangle.
\end{align}
Employing the correlation function (\ref{CBccgg1}), we can derive
the correlator
\begin{align}
\label{AcorrExtra3} \langle Y_{-2} B K c \Omega^{t_1} \gamma K
\Omega^{t_2} c \gamma \rangle = \frac{\pi  t_2
\left(t_1+t_2\right) \sin \left(\frac{\pi
   t_2}{t_1+t_2}\right)+\left(t_1^2+\left(2+\pi ^2\right) t_2 t_1+t_2^2\right) \cos
   \left(\frac{\pi  t_2}{t_1+t_2}\right)}{2 \pi ^2 \left(t_1+t_2\right){}^2}.
\end{align}
Plugging (\ref{AcorrExtra3}) into equation (\ref{AcorrExtra2}) and
performing the change of variables $t_1 \rightarrow  u v$, $t_2
\rightarrow  u - u v$, $\int_{0}^{\infty} dt_1 dt_2 \rightarrow
\int_{0}^{\infty} du \int_{0}^{1} dv \, u  $, the integral over
the variable $v$ can be easily done, so that we get
\begin{align}
\label{AcorrExtra4}  \mathcal{I}(\lambda) = \int_{0}^{\infty} du
\frac{\alpha_2 u e^{r_1 u} \big(\beta _2+\frac{u^2}{\pi^2}\alpha_2
(r_1-r_2)^2 + \alpha_2 \big)+\beta _2 u e^{r_2 u} \big(\alpha
_2+\frac{u^2}{\pi^2}\beta_2(r_1-r_2)^2 + \beta_2\big)}{ 2\pi^2 +
2(r_1-r_2)^2 u^2}.
\end{align}
The above integral exists only when $\Re \, r_{1,2} < 0$, and unlike the integral (\ref{AcorrA3}), here we were not able to write a simple analytic
expression for the result of this integral (\ref{AcorrExtra4}). Nevertheless, for the parameter $\lambda$ belonging to the region (\ref{condilam2}),
integrals like (\ref{AcorrExtra4}) can be evaluated numerically with arbitrary precision. The numerical evaluation of these type of integrals blows up in
the range where $\lambda \in (0,\kappa)$.

Carrying out similar computations for the rest of terms on the
R.H.S. of equation (\ref{NorV1H1Sollam1}), adding the results up
and performing numerical integration\footnote{The explicit
expression for the result of the energy in terms of integrals over
the variable $u$ is shown in appendix A.} together with the
definitions (\ref{defpqw})-(\ref{defr2}), (\ref{alphabeta1}),
(\ref{alphabeta2}), the energy turns out to be
\begin{align}
\label{AcorrA4} E(\widehat{\Phi}_{\lambda}) = \frac{\pi^2}{3} \Big[ -0.303963550927... \Big] = \frac{\pi^2}{3} \Big[ - \frac{3}{\pi^2} \Big] = - 1.
\end{align}

Collecting the results (\ref{NorV1H2A6}) and (\ref{AcorrA4}), we
can summarize the main result of our paper
\begin{align}
\label{FinalResult1} E(\widehat{\Phi}_{\lambda}) = \begin{cases}
      0, & \lambda = 0 \;,\;\; \text{Perturbative Vacuum Solution,} \\
      -1/2, & \lambda = 1/2 \;,\;\; \text{Half Brane Solution,} \\
      -1, & \big(\lambda < 0\big) \vee \big(\kappa \leq
\lambda <\frac{1}{2}\big) \vee \big(\lambda >\frac{1}{2}\big)
\;,\;\; \text{Tachyon Vacuum Solution,}
   \end{cases}
\end{align}
namely, depending on the value of the parameter $\lambda$, the
solution represents three distinct gauge orbits corresponding to
the perturbative vacuum, the half brane and the tachyon vacuum
solution.

\section{Summary and discussion}
We have studied and constructed a one-parameter family of solutions which contains the perturbative vacuum, the half brane and the tachyon vacuum solution
in the modified cubic superstring field theory. To our knowledge, this is the first explicit example of a solution which describes these three distinct
gauge orbits.

To evaluate the energy associated to the one-parameter family of
solutions we have performed analytic computations, however it
would be nice to confirm our results by employing numerical
techniques such as the curly $\mathcal{L}_0$ level expansion
\cite{Arroyo:2009ec,Arroyo:2011zt,AldoArroyo:2009hf} or the usual
Virasoro $L_0$ level expansion scheme
\cite{Moeller:2000xv,Kishimoto:2011zza,Arroyo:2014pua}. The
numerical analysis should be important, for instance, to check if
the solution behaves as a regular element in the state space
constructed out of Fock states
\cite{Schnabl:2010tb,Takahashi:2007du,AldoArroyo:2011gx}.

In the case of open bosonic string field theory, using elements of the $KBc$ subalgebra, in reference \cite{Erler:2012dz}, the existence of physically
distinct solutions has been analyzed such as the perturbative vacuum, the tachyon vacuum and the MNT ghost brane \cite{Okuda:2006fb,Masuda:2012kt}.
Following the lines developed in this paper, it would be nice to find a one-parameter family of solutions which describes these distinct gauge orbits.

Finally, we would like to comment that the construction of solutions based on gauge transformation of identity based solutions can be generalized in order
to consider more cumbersome solutions, such as the multibrane solutions \cite{AldoArroyo:2012if,Arroyo:2013pha}, and the recently proposed Erler's analytic
solution for tachyon condensation in Berkovits open superstring field theory \cite{Erler:2013wda}. Since the algebraic structure of Berkovits theory
\cite{Berkovits:1995ab} is similar to the cubic superstring field theory, the results of our work can be naturally extended, however the presence of a
non-polynomial action will bring us challenges in the search of new solutions within Berkovits theory.

\section*{Acknowledgements}
I would like to thank Ted Erler for useful discussions. I also
thank the referee for his appreciable work in the peer-review
process; his comments helped me to improve the paper. Finally, I
would like to give a special thank to my family: my wife Diany, my
son Davi, and my newborn daughter Sofia for their valuable company
during the elaboration of this work. This work has been supported
by CNPq grant 303073/2012-8.

\appendix
\setcounter{equation}{0}
\def\thesection{\Alph{section}}
\renewcommand{\theequation}{\Alph{section}.\arabic{equation}}

\section{Explicit expression for the energy of the one-parameter family of solutions}
Here we are going to write the explicit expression for the energy of the one-parameter family of solutions derived from the evaluation of all terms on the
R.H.S. of equation (\ref{NorV1H1Sollam1}). The result reads as follows
\begin{align}
\label{Apen1} E(\widehat{\Phi}_{\lambda}) = \frac{\pi^2}{3} \Big[
p^2 \mathcal{I}_1(\lambda) + pq \mathcal{I}_2(\lambda) + q^2
\mathcal{I}_3(\lambda) \Big],
\end{align}
where the functions $\mathcal{I}_1(\lambda)$,
$\mathcal{I}_2(\lambda)$ and $\mathcal{I}_3(\lambda)$ are given by
\begin{align}
\label{Apen2} \mathcal{I}_1(\lambda) &= \frac{-32 \lambda ^3+64
\lambda
^2-38 \lambda +3}{\pi ^2 \lambda ^2 (2 \lambda -1)^3}  \nonumber \\
&+\int_{0}^{\infty} du \frac{2 u^2 \Big[\alpha _2 e^{r_1 u} \big( \beta _2+ \frac{u^2}{\pi^2}\alpha _2(r_1-r_2)^2+\alpha _2\big)+\beta _2 e^{r_2 u} \big(
\alpha _2+ \frac{u^2}{\pi^2}\beta _2(r_1-r_2)^2 +\beta _2\big)\Big]}{\pi^2 + (r_1-r_2)^2 u^2}.
\end{align}

\begin{align}
\label{Apen3} \mathcal{I}_2(\lambda) = \int_{0}^{\infty} du \; \frac{u}{\pi ^2 \left(\pi ^2+u^2 \left(r_1-r_2\right){}^2\right){}^2 \left(r_1-r_2\right)} \Big[ \nonumber\\
\left(-e^{u r_1} \left(\alpha _2 \left(3 u r_2 \left(\pi ^2+u^2 r_2^2\right){}^2 \alpha _1+\left(\pi ^4+3 u^4 r_2^4\right) \beta _1\right)\right.\right.\nonumber\\
\left.+\left(\pi ^4+u r_2 \left(4 \pi ^4+u r_2 \left(4 \pi ^2+u r_2 \left(4 \pi ^2-u r_2\right)\right)\right)\right) \alpha _1 \beta _2\right)\nonumber\\
+e^{u r_2} \left(\left(\pi ^4-u r_2 \left(4 \pi ^4+u r_2 \left(-4
\pi ^2+u r_2 \left(6 \pi ^2+u r_2 \left(1+2 u
r_2\right)\right)\right)\right)\right)
\alpha _2 \beta _1\right.\nonumber\\
\left.+\left(\left(\pi ^4+u^3 r_2^3 \left(2 \pi ^2+u r_2 \left(3+2
u r_2\right)\right)\right) \alpha _1-3 u r_2 \left(\pi ^2+u^2
r_2^2\right){}^2
\beta _1\right) \beta _2\right)\nonumber\\
+u^4 r_1^4 \left(e^{u r_1} \left(\alpha _2 \left(-3 \beta _1+u r_2
\left(-15 \alpha _1+8 \beta _1\right)\right)+\left(1-8 u
r_2\right) \alpha _1 \beta
_2\right)\right.\nonumber\\
\left.-e^{u r_2} \left(\left(1+2 u r_2\right) \alpha _2 \beta _1+\left(-\left(3+2 u r_2\right) \alpha _1+15 u r_2 \beta _1\right) \beta _2\right)\right)\nonumber\\
+2 u^2 r_1^2 \left(-e^{u r_1} \left(u r_2 \alpha _2 \left(3
\left(3 \pi ^2+5 u^2 r_2^2\right) \alpha _1+\left(-2 \pi ^2+u r_2
\left(9-4 u r_2\right)\right)
\beta _1\right)\right.\right.\nonumber\\
\left.+\left(2 \pi ^2+u r_2 \left(8 \pi ^2+u r_2 \left(-3+4 u r_2\right)\right)\right) \alpha _1 \beta _2\right)\nonumber\\
+e^{u r_2} \left(-\left(-2 \pi ^2+u r_2 \left(7 \pi ^2+3 u r_2 \left(1+2 u r_2\right)\right)\right) \alpha _2 \beta _1\right.\nonumber\\
\left.\left.+u r_2 \left(\left(\pi ^2+3 u r_2 \left(3+2 u r_2\right)\right) \alpha _1-3 \left(3 \pi ^2+5 u^2 r_2^2\right) \beta _1\right) \beta _2\right)\right)\nonumber\\
+u^5 r_1^5 \left(3 e^{u r_2} \beta _1 \beta _2+e^{u r_1} \left(-2 \alpha _2 \beta _1+\alpha _1 \left(3 \alpha _2+2 \beta _2\right)\right)\right)\nonumber\\
+2 u^3 r_1^3 \left(e^{u r_2} \left(2 \left(\pi ^2+u r_2 \left(1+2
u r_2\right)\right) \alpha _2 \beta _1+\left(-2 u r_2 \left(3+2 u
r_2\right) \alpha_1\right.\right.\right.\nonumber\\
\left.\left.+3 \left(\pi ^2+5 u^2 r_2^2\right) \beta _1\right)
\beta _2\right)+e^{u r_1} \left(-\left(\pi ^2+6 u r_2 \left(-1+u
r_2\right)\right)
\alpha _2 \beta _1\right.\nonumber\\
\left.\left.+\alpha _1 \left(3 \left(\pi ^2+5 u^2 r_2^2\right) \alpha _2+\left(3 \pi ^2+2 u r_2 \left(-1+3 u r_2\right)\right) \beta _2\right)\right)\right)\nonumber\\
+u r_1 \left(e^{u r_2} \left(4 \left(\pi ^4+u r_2 \left(-2 \pi
^2+u r_2 \left(4 \pi ^2+u r_2 \left(1+2 u
r_2\right)\right)\right)\right) \alpha _2
\beta _1\right.\right.\nonumber\\
\left.+\left(-4 u^2 r_2^2 \left(\pi ^2+u r_2 \left(3+2 u
r_2\right)\right) \alpha _1+3 \left(\pi ^4+6 \pi ^2 u^2 r_2^2+5
u^4 r_2^4\right) \beta _1\right)
\beta _2\right)\nonumber\\
+e^{u r_1} \left(-2 u^2 r_2^2 \left(\pi ^2+u r_2 \left(-6+u
r_2\right)\right) \alpha _2 \beta _1+\alpha _1 \left(3 \left(\pi
^4+6 \pi ^2 u^2 r_2^2+5
u^4 r_2^4\right) \alpha _2 \right.\right.\nonumber\\
\left.\left.\left.\left.+2 \left(2 \pi ^4+u r_2 \left(4 \pi ^2+u
r_2 \left(7 \pi ^2+u r_2 \left(-2+u
r_2\right)\right)\right)\right) \beta
_2\right)\right)\right)\right)\Big].
\end{align}

\begin{align}
\label{Apen4} \mathcal{I}_3(\lambda) = \int_{0}^{\infty}du \;
\frac{u}{2 \pi ^4 \left(\pi ^2+u^2
\left(r_1-r_2\right){}^2\right){}^2 \left(r_1-r_2\right)}
\Big[\;\;\; \nonumber \\
\left(e^{u r_1} \left(8+\pi ^2\right) u^6
r_1^7 \alpha _2^2+e^{u r_1} \pi ^2 \left(-3 r_2 \left(\pi ^2+u^2
r_2^2\right){}^2 \alpha _2^2-2 \left(\pi
^4+3 u^4 r_2^4\right) \alpha _1 \beta _1\right)\right. \nonumber\\
+e^{u r_1} u^5 r_1^6 \left(\left(-8+\pi ^2\right) u \alpha _1^2+\alpha _2 \left(\left(9 \pi ^2-5 \left(8+\pi ^2\right) u r_2\right) \alpha _2+6 \pi
^2 \beta _2\right)\right)\nonumber\\
+u r_1^2 \left(e^{u r_1} \left(-8+\pi ^2\right) u \left(\pi ^4+6 \pi ^2 u^2 r_2^2+5 u^4 r_2^4\right) \alpha _1^2-e^{u r_1} \right.\nonumber\\
\left(-9 \pi ^6+u r_2 \left(\pi ^4 \left(26+\pi ^2\right)+u r_2 \left(-54 \pi ^4+u r_2 \left(2 \pi ^2 \left(23+\pi ^2\right)+u r_2 \left(-45 \pi
^2+\left(8+\pi ^2\right) u r_2\right)\right)\right)\right)\right) \nonumber\\
\alpha _2^2+4 \pi ^2 u \left(-2 e^{u r_1} \pi ^2+u^2 r_2^2 \left(3
e^{u r_2} \left(1+u r_2\right)+e^{u r_1} \left(-3+2 u
r_2\right)\right)\right)
\alpha _1 \beta _1\nonumber\\
+2 \pi ^2 \left(e^{u r_1} \left(\pi ^2+3 u^2 r_2^2\right) \left(3 \pi ^2+u r_2 \left(-8+u r_2\right)\right)-2 e^{u r_2} u r_2 \left(\pi ^2+u r_2 \left(5
\pi ^2+3 u r_2 \left(5+3 u r_2\right)\right)\right)\right) \nonumber\\
\left.\alpha _2 \beta _2-2 e^{u r_2} u r_2 \left(3 \pi ^2+5 u^2
r_2^2\right) \left(\left(-8+\pi ^2\right) u^2 r_2 \beta
_1^2+\left(3 \pi ^2+u r_2
\left(9 \pi ^2+\left(8+\pi ^2\right) u r_2\right)\right) \beta _2^2\right)\right)\nonumber\\
-e^{u r_2} \left(2 \pi ^2 \left(-\pi ^4+u r_2 \left(\pi ^4-4 \pi ^2 u r_2+u^3 r_2^3-u^4 r_2^4\right)\right) \alpha _1 \beta _1+r_2 \left(\pi ^2+u^2
r_2^2\right){}^2 \right.\nonumber\\
\left.\left(\left(-8+\pi ^2\right) u^2 r_2 \beta _1^2+6 \pi ^2
\left(1+u r_2\right) \alpha _2 \beta _2+\left(3 \pi ^2+u r_2
\left(9 \pi ^2+\left(8+\pi
^2\right) u r_2\right)\right) \beta _2^2\right)\right)\nonumber\\
+u^3 r_1^4 \left(2 e^{u r_1} \left(-8+\pi ^2\right) u \left(\pi ^2+5 u^2 r_2^2\right) \alpha _1^2+2 \pi ^2 u \left(e^{u r_2} \left(3+u r_2\right)+e^{u
r_1} \left(1+4 u r_2\right)\right) \alpha _1 \beta _1\right.\nonumber\\
-e^{u r_1} \alpha _2 \left(\left(-18 \pi ^4+u r_2 \left(63 \pi ^2+6 \pi ^4+10 u r_2 \left(-9 \pi ^2+\left(8+\pi ^2\right) u r_2\right)\right)\right)
\alpha _2\right.\nonumber\\
\left.-4 \pi ^2 \left(3 \pi ^2+u r_2 \left(-8+9 u r_2\right)\right) \beta _2\right)\nonumber\\
\left.-e^{u r_2} u r_2 \left(5 \left(-8+\pi ^2\right) u^2 r_2 \beta _1^2+2 \pi ^2 \left(7+3 u r_2\right) \alpha _2 \beta _2+5 \left(3 \pi ^2+u r_2
\left(9 \pi ^2+\left(8+\pi ^2\right) u r_2\right)\right) \beta _2^2\right)\right)\nonumber\\
+u^4 r_1^5 \left(e^{u r_2} \left(\left(-8+\pi ^2\right) u^2 r_2 \beta _1^2+\left(3 \pi ^2+u r_2 \left(9 \pi ^2+\left(8+\pi ^2\right) u r_2\right)\right)
\beta _2^2\right)+e^{u r_1} \left(10 \left(8+\pi ^2\right) u^2 r_2^2 \alpha _2^2\right.\right.\nonumber\\
\left.\left.+\pi ^2 \left(\left(19+2 \pi ^2\right) \alpha _2^2-2 u \alpha _1 \beta _1+6 \alpha _2 \beta _2\right)-u r_2 \left(5 \left(-8+\pi ^2\right)
u \alpha _1^2+3 \pi ^2 \alpha _2 \left(15 \alpha _2+8 \beta _2\right)\right)\right)\right)\nonumber\\
+u^2 r_1^3 \left(5 \left(8+\pi ^2\right) u^4 r_2^4 \left(e^{u r_1} \alpha _2^2+2 e^{u r_2} \beta _2^2\right)+\pi ^4 \left(6 e^{u r_2} \beta _2^2+e^{u
r_1} \alpha _2 \left(\left(14+\pi ^2\right) \alpha _2+12 \beta _2\right)\right)\right.\nonumber\\
+2 \pi ^2 u r_2 \left(e^{u r_2} \left(-8 u \alpha _1 \beta _1+\left(-8+\pi ^2\right) u \beta _1^2+\pi ^2 \beta _2 \left(2 \alpha _2+9 \beta _2\right)\right)\right.\nonumber\\
\left.-e^{u r_1} \left(3 \left(-8+\pi ^2\right) u \alpha _1^2+\pi ^2 \alpha _2 \left(27 \alpha _2+14 \beta _2\right)\right)\right)+2 u^3 r_2^3 \left(-e^{u
r_1} \right.\nonumber\\
\left.\left(5 \left(-8+\pi ^2\right) u \alpha _1^2+3 \pi ^2 \alpha
_2 \left(15 \alpha _2+4 \beta _2\right)\right)+e^{u r_2} \left(5
\left(-8+\pi
^2\right) u \beta _1^2+3 \pi ^2 \beta _2 \left(4 \alpha _2+15 \beta _2\right)\right)\right)+2 \pi ^2 \nonumber\\
\left.u^2 r_2^2 \left(3 e^{u r_1} \left(\left(13+\pi ^2\right)
\alpha _2^2-2 u \alpha _1 \beta _1+10 \alpha _2 \beta
_2\right)+e^{u r_2} \left(-4
u \alpha _1 \beta _1+\beta _2 \left(24 \alpha _2+\left(23+\pi ^2\right) \beta _2\right)\right)\right)\right)\nonumber\\
+r_1 \left(5 e^{u r_2} \left(8+\pi ^2\right) u^6 r_2^6 \beta _2^2+\pi ^4 u^2 r_2^2 \left(2 e^{u r_1} \alpha _2 \left(9 \alpha _2+2 \beta _2\right)+e^{u
r_2} \beta _2 \left(16 \alpha _2+\left(26+\pi ^2\right) \beta _2\right)\right)\right.\nonumber\\
+\pi ^6 \left(3 e^{u r_2} \beta _2^2+e^{u r_1} \left(2 u \alpha _1 \beta _1+3 \alpha _2 \left(\alpha _2+2 \beta _2\right)\right)\right)\nonumber\\
+\pi ^4 u r_2 \left(-e^{u r_1} \left(\left(-8+\pi ^2\right) u \alpha _1^2-8 u \alpha _1 \beta _1+\pi ^2 \alpha _2 \left(9 \alpha _2+4 \beta _2\right)\right)\right.\nonumber\\
\left.+e^{u r_2} \left(-8 u \alpha _1 \beta _1+\left(-8+\pi ^2\right) u \beta _1^2+\pi ^2 \beta _2 \left(4 \alpha _2+9 \beta _2\right)\right)\right)\nonumber\\
+u^5 r_2^5 \left(-e^{u r_1} \left(\left(-8+\pi ^2\right) u \alpha _1^2+9 \pi ^2 \alpha _2^2\right)+e^{u r_2} \left(5 \left(-8+\pi ^2\right) u \beta
_1^2+3 \pi ^2 \beta _2 \left(8 \alpha _2+15 \beta _2\right)\right)\right)\nonumber\\
+2 \pi ^2 u^3 r_2^3 \left(-e^{u r_1} \left(\left(-8+\pi ^2\right) u \alpha _1^2-8 u \alpha _1 \beta _1+\pi ^2 \alpha _2 \left(9 \alpha _2+2 \beta
_2\right)\right)\right.\nonumber\\
\left.+e^{u r_2} \left(3 \left(-8+\pi ^2\right) u \beta _1^2+\pi ^2 \beta _2 \left(14 \alpha _2+27 \beta _2\right)\right)\right)\nonumber\\
\left.\left.+\pi ^2 u^4 r_2^4 \left(e^{u r_1} \left(15 \alpha _2^2-2 u \alpha _1 \beta _1+14 \alpha _2 \beta _2\right)+e^{u r_2} \left(-8 u \alpha _1 \beta
_1+\beta _2 \left(32 \alpha _2+3 \left(21+2 \pi ^2\right) \beta _2\right)\right)\right)\right)\right)\Big].
\end{align}

The above integrals converge provided that $\Re \, r_{1,2} < 0$,
and can be computed numerically with arbitrary precision.



\begin{thebibliography}
\bibitem{Schnabl:2005gv}
  M.~Schnabl, \textit{Analytic solution for tachyon condensation in open string field
theory},
  Adv.\ Theor.\ Math.\ Phys.\  {\bf 10}, 433 (2006)
  [hep-th/0511286].

\bibitem{Erler:2009uj}
  T.~Erler and M.~Schnabl, \textit{A Simple Analytic Solution for Tachyon Condensation},
  JHEP {\bf 0910}, 066 (2009)
  [arXiv:0906.0979 [hep-th]].

\bibitem{Okawa:2006vm}
  Y.~Okawa, \textit{Comments on Schnabl's analytic solution for tachyon condensation
in Witten's open string field theory},
  JHEP {\bf 0604}, 055 (2006)
  [hep-th/0603159].

\bibitem{Witten:1985cc}
  E.~Witten, \textit{Noncommutative Geometry and String Field Theory},
  Nucl.\ Phys.\ B {\bf 268}, 253 (1986).

\bibitem{Erler:2007xt}
  T.~Erler, \textit{Tachyon Vacuum in Cubic Superstring Field Theory},
  JHEP {\bf 0801}, 013 (2008)
  [arXiv:0707.4591 [hep-th]].

\bibitem{Aref'eva:2008ad}
  I.~Y.~Aref'eva, R.~V.~Gorbachev and P.~B.~Medvedev, \textit{Tachyon Solution in Cubic Neveu-Schwarz String Field Theory},
  Theor.\ Math.\ Phys.\  {\bf 158}, 320 (2009)
  [arXiv:0804.2017 [hep-th]].

\bibitem{Gorbachev:2010zz}
  R.~V.~Gorbachev, \textit{New solution of the superstring equation of motion},
  Theor.\ Math.\ Phys.\  {\bf 162}, 90 (2010)
  [Teor.\ Mat.\ Fiz.\  {\bf 162}, 106 (2010)].

\bibitem{Arefeva:1989cp}
  I.~Y.~Arefeva, P.~B.~Medvedev and A.~P.~Zubarev, \textit{New Representation for String Field Solves the Consistency Problem
for Open Superstring Field Theory},
  Nucl.\ Phys.\ B {\bf 341}, 464 (1990).

\bibitem{Arroyo:2010fq}
  E.~A.~Arroyo, \textit{Generating Erler-Schnabl-type Solution for Tachyon Vacuum in Cubic Superstring Field Theory},
  J.\ Phys.\ A {\bf 43}, 445403 (2010)
  [arXiv:1004.3030 [hep-th]].

\bibitem{Zeze:2010sr}
  S.~Zeze, \textit{Regularization of identity based solution in string field theory},
  JHEP {\bf 1010}, 070 (2010)
  [arXiv:1008.1104 [hep-th]].

\bibitem{Arroyo:2010sy}
  E.~A.~Arroyo, \textit{Comments on regularization of identity based solutions in string field theory},
  JHEP {\bf 1011}, 135 (2010)
  [arXiv:1009.0198 [hep-th]].

\bibitem{Arefeva:2010yd}
  I.~Y.~Arefeva and R.~V.~Gorbachev, \textit{On Gauge Equivalence of Tachyon Solutions in Cubic Neveu-Schwarz
String Field Theory},
  Theor.\ Math.\ Phys.\  {\bf 165} (2010) 1512
  [arXiv:1004.5064 [hep-th]].

\bibitem{Erler:2012qn}
  T.~Erler and C.~Maccaferri, \textit{Connecting Solutions in Open String Field Theory with Singular
Gauge Transformations},
  JHEP {\bf 1204}, 107 (2012)
  [arXiv:1201.5119 [hep-th]].

\bibitem{Kishimoto:2014lua}
  I.~Kishimoto, T.~Masuda and T.~Takahashi, \textit{Observables for identity-based tachyon vacuum solutions},
  PTEP {\bf 2014}, no. 10, 103B02 (2014)
  [arXiv:1408.6318 [hep-th]].

\bibitem{Kishimoto:2009nd}
  I.~Kishimoto and T.~Takahashi, \textit{Vacuum structure around identity based solutions},
  Prog.\ Theor.\ Phys.\  {\bf 122}, 385 (2009)
  [arXiv:0904.1095 [hep-th]].

\bibitem{Inatomi:2011an}
  S.~Inatomi, I.~Kishimoto and T.~Takahashi, \textit{Homotopy Operators and Identity-Based Solutions in Cubic
Superstring Field Theory},
  JHEP {\bf 1110}, 114 (2011)
  [arXiv:1109.2406 [hep-th]].

\bibitem{Erler:2012dz}
  T.~Erler,\textit{ The Identity String Field and the Sliver Frame Level Expansion},
  JHEP {\bf 1211}, 150 (2012)
  [arXiv:1208.6287 [hep-th]].

\bibitem{Erler:2010pr}
  T.~Erler, \textit{Exotic Universal Solutions in Cubic Superstring Field Theory},
  JHEP {\bf 1104}, 107 (2011)
  [arXiv:1009.1865 [hep-th]].

\bibitem{AldoArroyo:2012if}
  E.~Aldo Arroyo, \textit{Multibrane solutions in cubic superstring field theory},
  JHEP {\bf 1206}, 157 (2012)
  [arXiv:1204.0213 [hep-th]].

\bibitem{Arroyo:2013pha}
  E.~Aldo Arroyo, \textit{Comments on multibrane solutions in cubic superstring field
theory},
  PTEP {\bf 2014}, no. 6, 063B03 (2014)
  [arXiv:1306.1865 [hep-th]].

\bibitem{Erler:2006hw}
  T.~Erler, \textit{Split String Formalism and the Closed String Vacuum},
  JHEP {\bf 0705}, 083 (2007)
  [hep-th/0611200].

\bibitem{Erler:2006ww}
  T.~Erler, \textit{Split String Formalism and the Closed String Vacuum, II},
  JHEP {\bf 0705}, 084 (2007)
  [hep-th/0612050].

\bibitem{Schnabl:2010tb}
  M.~Schnabl, \textit{Algebraic solutions in Open String Field Theory - A Lightning
Review},
  Acta Polytechnica 50, no. 3 (2010) 102
  [arXiv:1004.4858 [hep-th]].

\bibitem{Erler:2013wda}
  T.~Erler, \textit{Analytic solution for tachyon condensation in Berkovits` open
superstring field theory},
  JHEP {\bf 1311}, 007 (2013)
  [arXiv:1308.4400 [hep-th]].

\bibitem{Berkovits:1995ab}
  N.~Berkovits, \textit{SuperPoincare invariant superstring field theory},
  Nucl.\ Phys.\ B {\bf 450}, 90 (1995)
  [Nucl.\ Phys.\ B {\bf 459}, 439 (1996)]
  [hep-th/9503099].

\bibitem{Arefeva:1988nn}
  I.~Y.~Arefeva and P.~B.~Medvedev, \textit{Anomalies in Witten's Field Theory of the Nsr String},
  Phys.\ Lett.\ B {\bf 212}, 299 (1988).

\bibitem{Berkovits:2000hf}
  N.~Berkovits, A.~Sen and B.~Zwiebach, \textit{Tachyon condensation in superstring field theory},
  Nucl.\ Phys.\ B {\bf 587}, 147 (2000)
  [hep-th/0002211].

\bibitem{Ohmori:2003vq}
  K.~Ohmori,\textit{ Level expansion analysis in NS superstring field theory
revisited},
  hep-th/0305103.

\bibitem{Arefeva:2002mb}
  I.~Y.~Arefeva, D.~M.~Belov and A.~A.~Giryavets, \textit{Construction of the vacuum string field theory on a non-BPS
brane},
  JHEP {\bf 0209}, 050 (2002)
  [hep-th/0201197].

\bibitem{Arroyo:2009ec}
  E.~A.~Arroyo, \textit{Cubic interaction term for Schnabl's solution using Pade
approximants},
  J.\ Phys.\ A {\bf 42}, 375402 (2009)
  [arXiv:0905.2014 [hep-th]].

\bibitem{Arroyo:2011zt}
  E.~A.~Arroyo, \textit{Conservation laws and tachyon potentials in the sliver frame},
  JHEP {\bf 1106}, 033 (2011)
  [arXiv:1103.4830 [hep-th]].

\bibitem{AldoArroyo:2009hf}
  E.~Aldo Arroyo, \textit{The Tachyon Potential in the Sliver Frame},
  JHEP {\bf 0910}, 056 (2009)
  [arXiv:0907.4939 [hep-th]].

\bibitem{Moeller:2000xv}
  N.~Moeller and W.~Taylor, \textit{Level truncation and the tachyon in open bosonic string field
theory},
  Nucl.\ Phys.\ B {\bf 583}, 105 (2000)
  [hep-th/0002237].

\bibitem{Kishimoto:2011zza}
  I.~Kishimoto, \textit{On numerical solutions in open string field theory},
  Prog.\ Theor.\ Phys.\ Suppl.\  {\bf 188}, 155 (2011).

\bibitem{Arroyo:2014pua}
  E.~Aldo Arroyo, \textit{Level truncation analysis of a simple tachyon vacuum solution in
cubic superstring field theory},
  JHEP {\bf 1412}, 069 (2014)
  [arXiv:1409.1890 [hep-th]].

\bibitem{Takahashi:2007du}
  T.~Takahashi, \textit{Level truncation analysis of exact solutions in open string field
theory},
  JHEP {\bf 0801}, 001 (2008)
  [arXiv:0710.5358 [hep-th]].

\bibitem{AldoArroyo:2011gx}
  E.~Aldo Arroyo, \textit{Level truncation analysis of regularized identity based
solutions},
  JHEP {\bf 1111}, 079 (2011)
  [arXiv:1109.5354 [hep-th]].

\bibitem{Okuda:2006fb}
  T.~Okuda and T.~Takayanagi, \textit{Ghost D-branes},
  JHEP {\bf 0603}, 062 (2006)
  [hep-th/0601024].

\bibitem{Masuda:2012kt}
  T.~Masuda, T.~Noumi and D.~Takahashi, \textit{Constraints on a class of classical solutions in open string field
theory},
  JHEP {\bf 1210}, 113 (2012)
  [arXiv:1207.6220 [hep-th]].

\end{thebibliography}
\end{document}